\documentclass[prl,aps,twocolumn,floatfix,showpacs]{revtex4}
\usepackage{amsmath,graphics,epsfig,color,verbatim,ulem}

 \newcommand{\vk}{{\mathbf{k}}}

\begin{document}
\title{Correlated electronic structure of LaO$_{1-x}$F$_{x}$FeAs}

\author{K. Haule, J. H. Shim, and G. Kotliar}

\affiliation{Department of Physics, Rutgers University,
Piscataway, NJ 08854, USA}

\begin{abstract}
  We compute the electronic structure, momentum resolved spectral
  function and optical conductivity of the new superconductor
  LaO$_{1-x}$F$_x$FeAs within the combination of the Density
  functional theory and the Dynamical Mean Field Theory. We find that
  the compound in the normal state is a strongly correlated metal and
  the parent compound is a bad metal at the verge of the metal
  insulator transition. We argue that the superconductivity is not
  phonon mediated.
\end{abstract}
 \date{\today} 
\maketitle

In the Bardeen-Cooper-Schrieffer theory of superconductivity,
electrons form Cooper pairs through an interaction mediated by
vibrations of the crystal. Like lattice
vibrations, antiferromagnetic fluctuations can also produce an
attractive interaction creating Cooper pairs, though with spin and
angular momentum properties different from those of conventional
superconductors. Such interactions was implicated for class of heavy
fermion materials based on Ce, U and Pu with rather low transition
temperatures, and cuprate superconductors with the highest known
transition temperatures. Recently a surprising discovery of
superconductivity in iron-based compound LaO$_{1-x}$F$_x$FeP
\cite{LaOFeP} with $T_c\sim 7\,$K sparked a new direction to explore
superconductivity in a completely new class of materials. Very recently
a substitution of P by As raised T$_c$ to $26\,$K \cite{LaOFeAs}
becoming already one of the superconductors with highest $T_c$ among
non-cuprate based materials. Exploring superconductivity in similar
iron based compounds holds a lot of promise for increasing $T_c$.

In this letter we explore the electronic structure and optical
properties of LaO$_{1-x}$F$_x$FeAs within Density Functional Theory
(DFT) and Dynamical Mean Field Theory (DMFT).

\begin{figure}[!ht]
\centering{
  \includegraphics[width=0.4\linewidth]{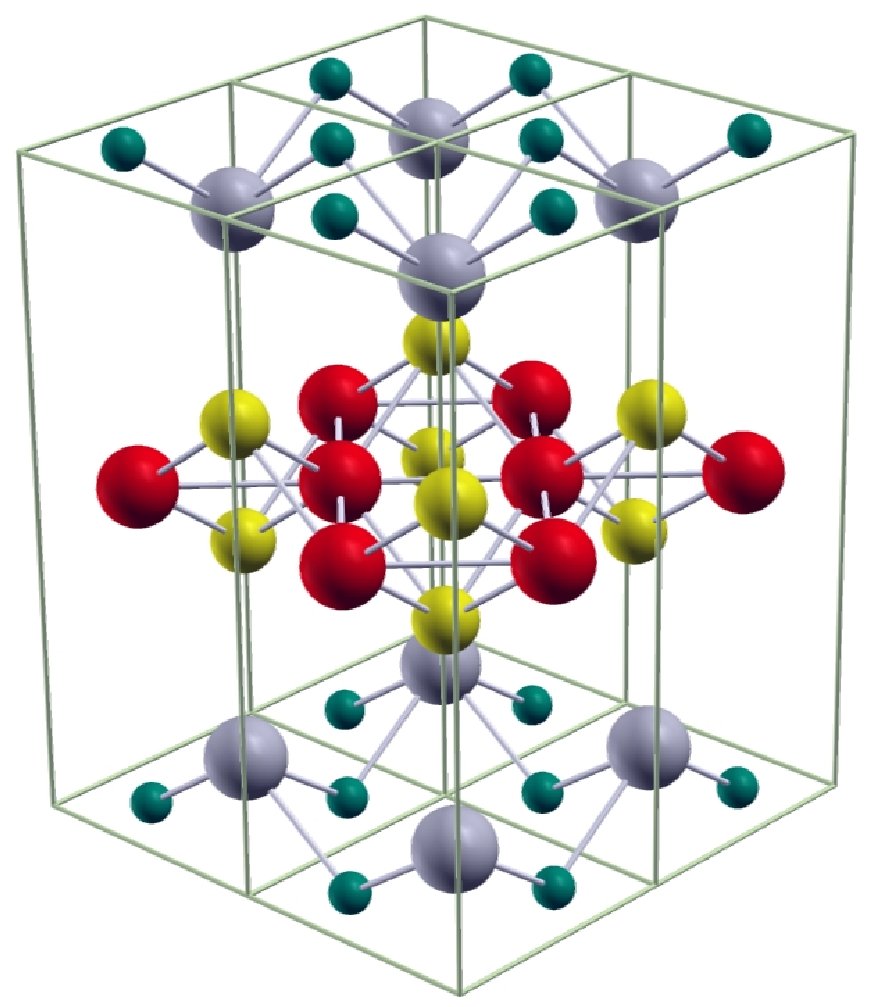}
  }
\caption{
Crystal structure of LaOFeAs. The middle layer consists of Fe atoms
(red spheres) which form a two dimensional square lattice and As atoms
(yellow). The second plane consist of O (green) and La (gray)
atoms. To describe the orbital character in Fig.~\ref{DMFT1}, the
coordinate system is chosen such that both $x$ and $y$ axis point
from each Fe atom towards the nearest neighbor Fe atoms.
}
\label{struct}
\end{figure}

LaOFeAs has a layered tetragonal crystal structure shown in
Fig.\ref{struct}.  Layers of La and O are sandwiched between layers
of Fe and As. The Fe atoms form a square two dimensional lattice with
Fe-Fe lattice spacing of 2.853$\textrm{\AA}$.
\begin{figure}[!ht]
\centering{
  \includegraphics[width=0.8\linewidth]{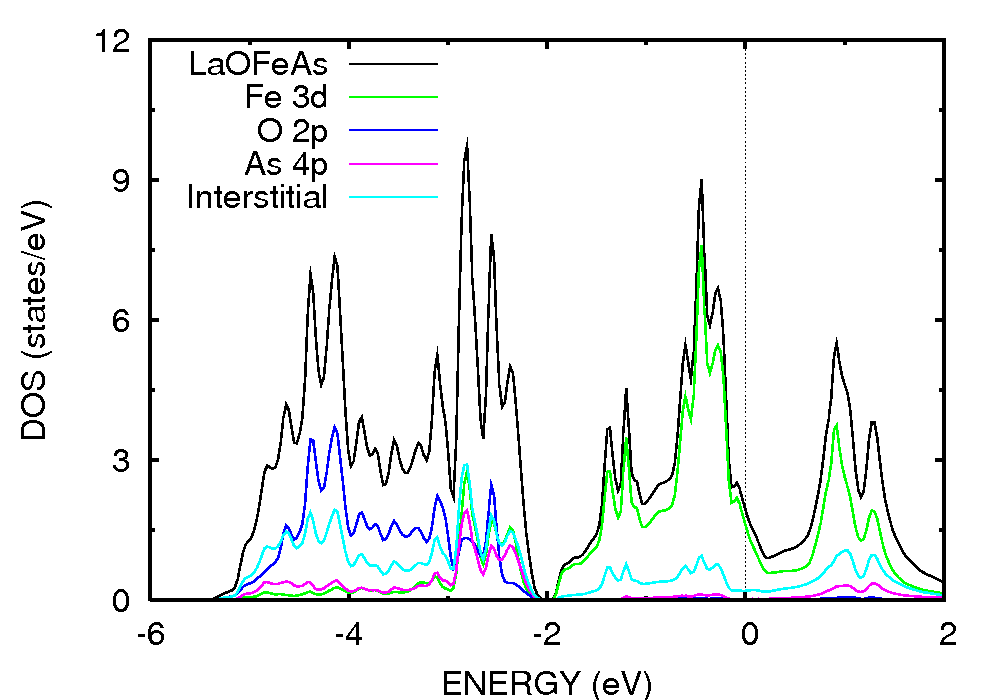}
  }
\caption{
Density of states within the GGA approximation. The partial
character of Fe, O and As is shown separately. 
Due to presence of As, a lot of electronic charge is found in the
interstitial regions and can not be assigned an atomic character.
}
\label{DOSWien}
\end{figure}
To understand the material properties, it is important to identify the
character of dominant bands near the Fermi level, their energy and
momentum distribution. For this purpose, the first principles density
functional theory is the invaluable tool. We used the full-potential
augmented plane-wave method as implemented in Wien code \cite{Wien}.
The lattice parameters and internal atomic positions have been
determine experimentally ($a=4.035$, $c=8.741$, $z_{La}=0.142$,
$z_{As}=0.651$).  For the exchange correlation potential we used
gradient approximation \cite{GGA} (GGA) in the Perdew-Burke-Ernzerhof
variant, and $12\times 12\times 5$ k-points.

This method predicts that the dominant states at the Fermi level come
from Fe $3d$ atomic states extending roughly between -2$\,$eV and
2$\,$eV as shown in Fig.~\ref{DOSWien}.

The important feature of LaOFeAs compound is that DFT predicts a very
steep and negative slope of the density of states (DOS) at the Fermi
level.  In the rigid band approximation, neglecting the many body
effects, the hole doping will lead to increases of the DOS at the
Fermi level, while the electron doping will be accompanied by a
decrease of DOS.  Hence, for the conventional phonon mediated
superconductor, one would expect decrease (increase) of $T_c$ when
doping electrons (holes), contrary to what is observed in experiment
\cite{LaOFeAs}. According to Ref.~\cite{LaOFeAs}, the electron doping
by $F^-$ leads to superconducting ground state, while $Ca^{2+}$ hole
doping does not cause superconductivity. This gives a clear hint that
the superconductivity in this compound might not be phonon mediated.
Indeed an explicit calculation of the phonon coupling constants within
the DFT, gives a value too small
to explain the observed critical temperature.

While the dominant electronic character near the Fermi level is due to
Fe, a strong mixing with As is apparent at $-2.7\,$eV where As $4p$
band is strongly peaked. There is a noticeable similarity between the
Fe-As mixing and oxygen-transition metal mixing in transition metal
oxides.  However, there is an important difference between the two.
The As bands are much more itinerant and broader than oxygen $2p$
bands.  Consequently, plenty of As electronic charge is found in the
interstitial region far from any of the constituting atoms, which is
shown separately in Fig.~\ref{DOSWien}.

While the crystal field splitting in most of transition metal oxides
leads to a clear separation of transition metal $d$ bands into
$t_{2g}$ and $e_g$ part, this is not the case in LaOFeAs due to the
itinerant character of As. All five Fe-$d$ bands therefore participate
in the bonding and in the Fermi surface of the compound.

Within DFT, the parent compound is very close to magnetic instability
and at normal pressure DFT selects ferromagnetic state while 10\%
compressed solid preferes paramagnetic state. The magnetic moment
predicted by DFT is very small of the order of 0.15$\mu_B$. We note in
passing that within DFT the antiferromagnetic state is unstable at
normal pressure.

The bandwidth of iron $3d$ bands is only around $3\,$eV while the
Coulomb repulsion in this orbital is around $4\,$eV
\cite{arsetivan}. 
For a single band system with this bandwidth and Coulomb repulsion,
one would expect a Mott insulating ground state. On the other hand,
the perfectly degenerate system with five degenerate $d$ bands would
still be below the Mott transition, since the critical $U$ for the
disappearance of the Fermi liquid solution within DMFT is proportional
to $N$, where $N$ is the band degeneracy of the system (in case of $d$
electrons 5). The five $d$ bands of Fe in LaOFeAs are not perfectly
degenerate, but one would still expect relatively large critical $U$
for the Mott transition for the odd number of electrons in the $d$
band. In the parent compound LaOFeAs, the number of $d$ electrons is
6. The Coulomb repulsion can strongly enhance the splitting between
the bands \cite{Millis} and the gap can open at the Fermi level. This
interplay of crystal field splitting and correlation effects was
addressed in many model Hamiltonian studies
\cite{Manini,Poteryaev,Millis} and it was shown how a bad metal or bad
semiconductor can appear on the metallic side at finite temperature.
In LaOFeAs we checked that a slightly enhanced Coulomb repulsion
($U=4.5\,$eV) leads to a finite gap in the $d$ band. For typical
Coulomb repulsion of Fe ($U=4\,$eV) the system is still metallic, but a
bad metal having, some characteristics of a bad semiconductor. The
metallic state is however still very correlated with quasiparticle
renormalization amplitude between $Z\sim 0.2-0.3$.

To describe this type of system, one needs to go beyond the
traditional band structure methods and concentrate on the spectral
function $A(\vk,\omega)_{LL'}$ of the system which described the
probability to add or remove an electron with momentum $\vk$, angular
momentum $L=(l,m)$ and energy $\omega$.

To compute the spectral function of the LaOFeAs, we used Dynamical
Mean Field Theory (DMFT) \cite{old-review,new-review} which takes
into account the strong Coulomb repulsion among the correlated set of
bands and can describe the dual nature of the electrons in correlated
compounds; the itinerant part of spectra which forms narrow
quasiparticle bands at low energy, and the localized part of spectra
at higher energy, which correspond to the nearly localized atomic
nature of the electron. 
In this method, the spectral function is obtained from the one electron Green's
function $A(\vk,\omega)=(G^\dagger(\vk,\omega)-G(\vk,\omega))/(2\pi i)$
where the latter takes the form
\begin{eqnarray}
G(\vk,\omega)=\frac{1}{O_\vk(\omega+\mu)-H_\vk-\Sigma(\omega)}.
\label{equation1}
\end{eqnarray}
The one electron part of the Hamiltonian $H_\vk$ and overlap matrix
$O_\vk$ is obtained by the LDA method \cite{Savrasov} while the
self-energy is computed by solving an auxiliary quantum impurity
problem embedded in a self-consistent medium.  We used the numerically
exact continuous time quantum Monte Carlo method \cite{my-PRB}. The
on-site coulomb repulsion on Fe-$3d$ bands is estimated to be $4\,$eV
\cite{arsetivan} and the Hunds coupling of $J=0.7\,$eV. The
temperature is fixed at $116\,$K.

\begin{figure}[!ht]
\centering{
  \includegraphics[width=0.8\linewidth]{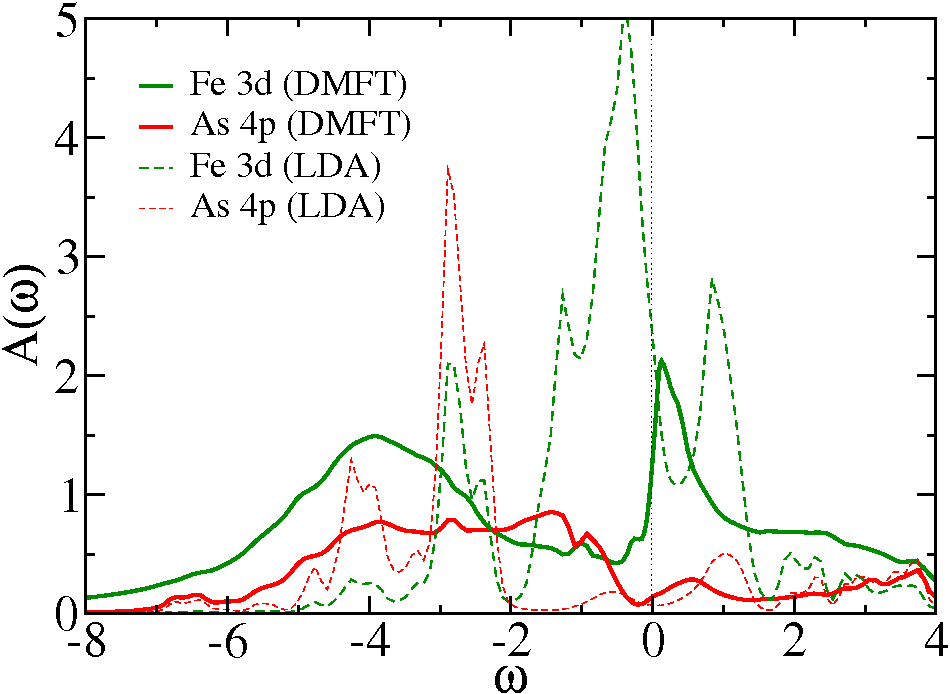}
  \includegraphics[width=0.8\linewidth]{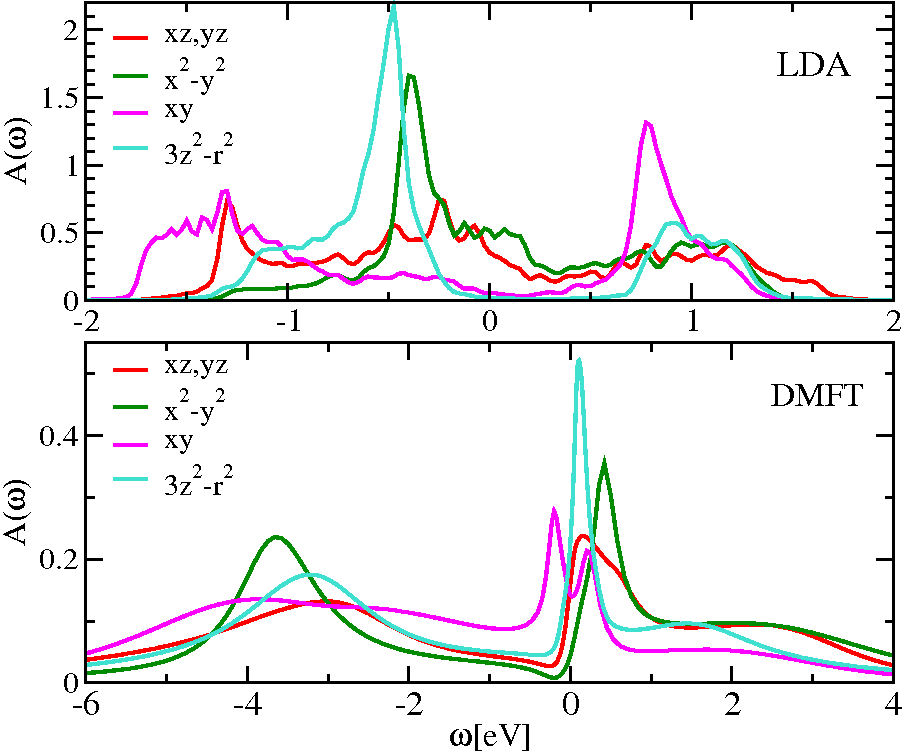}
  }
\caption{
a) DMFT density of states compared with LDA DOS.
b) Orbitally resolved Fe-$3d$ density of states within LDA and DMFT.
}
\label{DMFT1}
\end{figure}
The local spectral function $A(\omega)=\sum_\vk A(\vk,\omega)$ at
temperature $T=116K$ is shown in Fig.~\ref{DMFT1}a together with the
corresponding LDA density of states.  The DMFT approach predicts a
renormalized low energy band with a fraction of the original width
($Z\sim 0.2-0.3$) while most of the weight is transferred into a broad
Hubbard band at the binding energy $\sim -4\,$eV. The system remains
metallic at finite temperatures although very bad metal with the
scattering rate at the Fermi level as high as $0.4\,$eV at $116\,$K.
With decreasing temperature the Fermi surface is shrinking and the
semiconducting gap is likely to open at zero temperature. 
Indeed slightly enhanced Coulomb repulsion ($U=4.5\,$eV) opens the gap
even at room temperature. The correlation enhanced splitting between
different orbitals leads to separation of bands into those that act as
fully filled or fully empty bands at low energy. The high energy
Hubbard bands are only weakly effected by such splitting.  The Coulomb
repulsion thus strongly reduces carrier density and pushes the parent
compound on the verge of the transition between a bad metal and a bad
semiconductor.  At the same time, the localization of electrons leads
to local moment formation and enhancement of the spin susceptibility
in the doped compound.

\begin{figure}[!ht]
\centering{
  \includegraphics[width=0.9\linewidth]{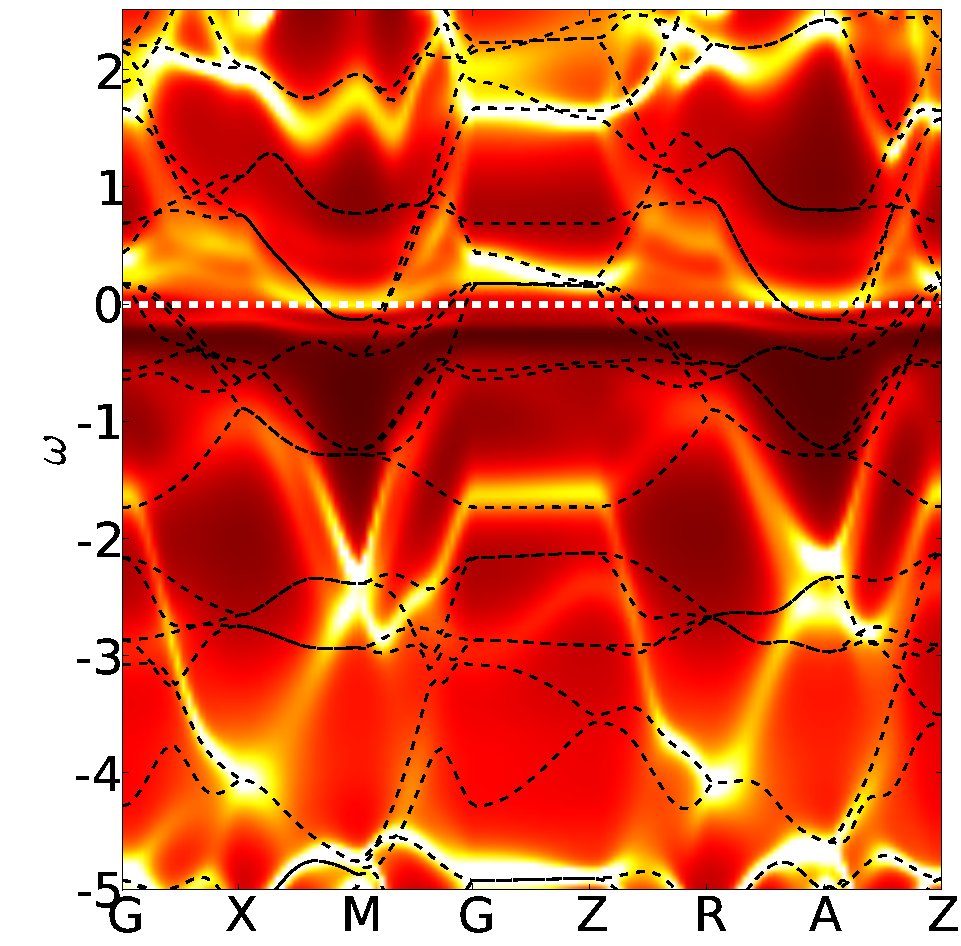}
  \includegraphics[width=0.9\linewidth]{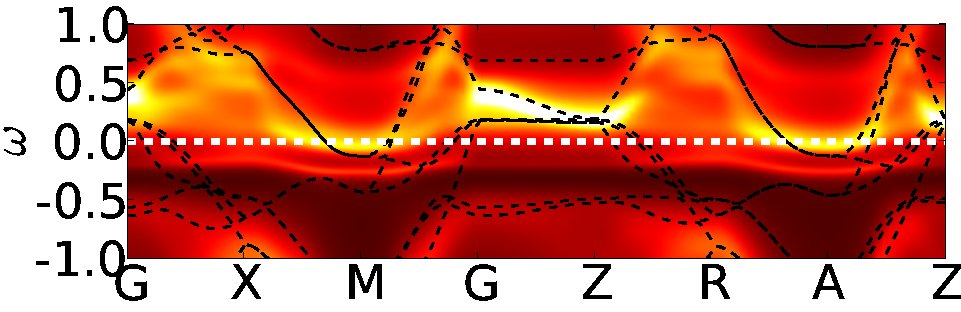}
  }
  \caption{ Momentum resolved spectral function $A_{4d}(\vk,\omega)$
    within DMFT (color coding) together with the LDA band structure
    (dashed lines).  The upper panel corresponds to the undoped parent
    compound while the lower panel shows the $10\%$ electron doped
    compound in the virtual crystal approximation.  }
\label{DMFT2}
\end{figure}

Many of the unconventional superconductors are known to have a very
simple low energy band structure. For example, in the copper oxides, a
single band is crossing the Fermi level. Similarly, the Fermi surface
of the Na doped cobaltates \cite{cobaltates} has primarily $a_{1g}$
single sheet Fermi surface. The situation is very different in LaOFeAs
within LDA. As can be seen in Fig.~\ref{DMFT1}b LDA predicts that all
five Fe $3d$ orbitals have finite weight close to or at the Fermi
level.  The situation is not simplified when the Coulomb correlation
is accounted for. The spectral weight splits into high energy
incoherent part and low energy part, which is very asymmetric and
considerably reduced due to proximity to the semiconducting state.
Upon doping, the quasiparticle peaks move to the Fermi level, the
scattering rate is reduced, and the system becomes better conductor.
Experimentally, doping leads to the superconducting ground state at
low temperature, which is likely to be of unconventional origin.  The
cooper pairs are likely to be formed out of composite singlets of spin
and orbital degrees of freedom.

Fig.~\ref{DMFT2}a shows momentum resolved spectral function
$\sum_{L}A(\vk,\omega)_{LL}$ in color coding together with the LDA
bands in the energy range between $-5\,$eV and $2.5\,$eV and momentum
dispersion in the high symmetry directions of the first Brillouin zone.
The higher energy band structure at $\omega>2\,$eV and $\omega<-5\,$eV
is not considerably different in the two approaches.  In the
intermediate frequency region between $-1.5\,$eV and $1.5\,$eV a
depletion of the spectral weight is apparent in DMFT approach.  While
LDA predicts a large number of bands in this region, DMFT redistributes
most of this weight further away from the Fermi level to the range
between $-2\,$eV and $-4\,$eV. Consequently a set of states with large
scattering rate and short lifetime is predicted by DMFT.  
Finally the low energy part of the spectra is considerably modified
when Coulomb correlation is taken into account. The hole pockets
around $\Gamma$ point are barely identifiable and the electron pockets
centered at $M$ and $A$ almost disappear - they are only touching the
Fermi level and the weigh below the Fermi level is only due to large scattering
rate. Thus the semiconducting type of gap is opening at the Fermi
level. The correlations thus enhance the crystal field splittings
which ultimately lead to a state being a band-semiconductor acting as
bad metal at finite temperatures.

The situation is different in the doped compound (see Fig.~\ref{DMFT2}b)
where the electron pockets clearly cross the Fermi level.  The band
velocity and effective mass are considerably enhanced ($3-5$-times)
while the scattering rate still remains large. Finally, the hole
pockets around $\Gamma$ remain highly scattered.

\begin{figure}[!ht]
\centering{
  \includegraphics[width=0.85\linewidth]{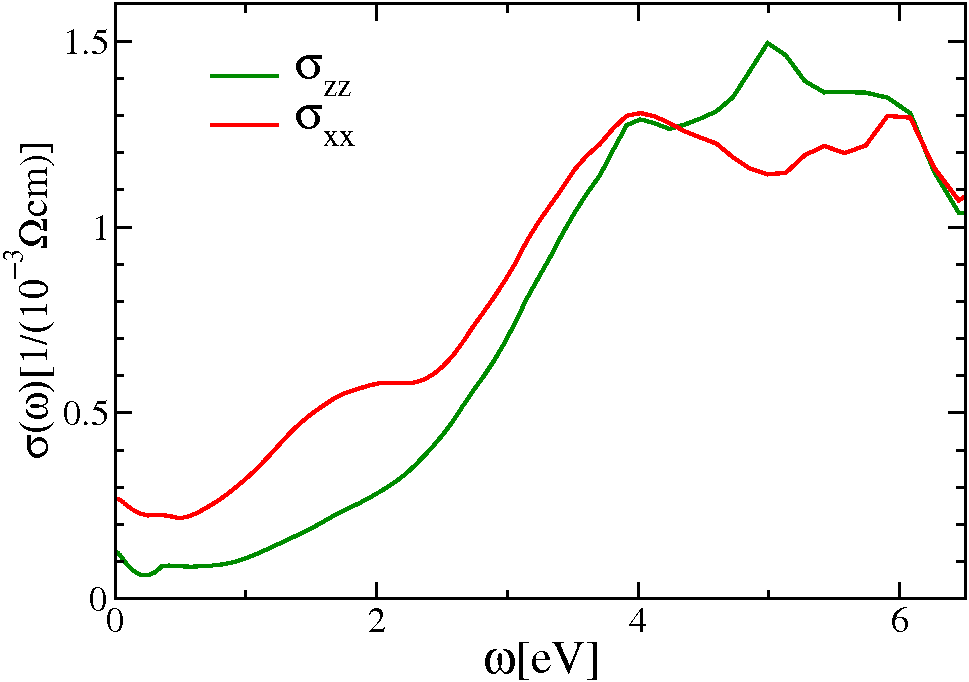}
  }
\caption{
Optical conductivity of LaOFeAs within DMFT: in-plane and out-of plane.
}
\label{DMFT3}
\end{figure}
We also compute optical conductivity of LaOFeAs within the DMFT in
both $xy$ plane and along the $z$ direction. 
Considerable anisotropy can be identified in Fig.~\ref{DMFT2}b.
The Drude peak, which is a hallmark of metallicity and Fermi liquid
state, is absent because of proximity to the semiconducting state. Our
calculations predict small number of carriers and large scattering
rate, due to partial localization of carriers in this compound, and
consequently low conductivity. One can identify a shoulder in optical
conductivity close to $2\,$eV. It corresponds to transitions between
Fe-$3d$ and As $4p$ states. The large peak at $4\,$eV is mostly due to
transitions within the $d$ shell from the lower Hubbard band to the
quasiparticle peak above the Fermi level.

In this work we used the single site Dynamical Mean Field Theory which
approximates the full momentum dependent dynamical self-energy by its
purely local component $\Sigma(\omega)$ (see Eq.~\ref{equation1})
which is still a $10\times 10$ matrix in orbital space of Fe-$d$
orbitals. While this approximation successfully describes physics of
many $f$ and $d$ electron systems, it does not capture the low energy
physics of cuprate superconductors where momentum space
differentiation leads to Fermi surface of arc-shape. The cluster
extensions of DMFT can capture some of the reach physics of cuprates.
We can not exclude the possibility that the low energy band structure
of doped LaOFeAs will also require similar non-local correlations, and
we leave this question open for future work.

Finally we comment on possible routes to understand the
superconductivity near a bad semiconductor in this compound.  Since
the Coulomb correlation is below the critical $U$ for this system to
undergo a Mott transition, an itinerant approach might be adequate.
This will involve a generalization of the spin-fluctuation exchange
mechanism to a multiorbital situation. In adition the pairing
interaction might involve not just spin but also orbital
fluctuations within the $d$ subshell.

In conclusion, we studied the band structure of newly discovered
superconductor LaO$_{1-x}$F$_x$FeAs, and we predict the orbital and
momentum resolved spectral function and optical conductivity of the
compound. Density functional theory predicts that a set of Fe $3d$
bands are crossing the Fermi level with no clear splitting into $e_g$
and $t_{2g}$ manifold. The coulomb correlations among the six
electrons in the set of five Fe-$3d$ orbitals is strong enough to push
the compound close to the metal insulator transition.  The correlation
enhance the crystal field splitting among the $d$ orbitals which in
turn increase correlations. At temperature $T=116\,$K studied here,
the parent compound LaOFeAs is still metallic although a bad metal
with very large scattering rate and strongly reduced number of
carriers. Doping the parent compound leads to electron pockets
centered at M and A point in momentum resolved spectral function.

\end{document}